**Bridge2AI: Building A Cross-disciplinary Curriculum Towards AI-Enhanced Biomedical and Clinical Care**

John Rincon, BS[1,#,*], Alexander R. Pelletier, PhD[1,*], Destiny Gilliland, BS[1], Wei Wang, PhD[1], Ding Wang, PhD[1], Baradwaj S. Sankar, BS[1], Lori Scott-Sheldon, PhD[2], Samson Gebreab, PhD[3], William Hersh, MD[4], Parisa Rashidi, PhD[5], Sally Baxter, MD[6], Wade Schulz, PhD[7], Trey Ideker, PhD[8], Yael Bensoussan, MD[9], Paul C. Boutros, PhD[1], Alex A.T. Bui, PhD[1], Colin Walsh, MD[10], Karol E. Watson, PhD[1], Peipei Ping, PhD[1,#]

1. Bridge Center of BRIDGE2AI at UCLA, Los Angeles, CA 90095, USA.
2. National Institute of Mental Health, NIH, Bethesda, MD 20892, USA.
3. Office of Data Science Strategy, NIH, Bethesda, MD 20892, USA.
4. Department of Medical Informatics & Clinical Epidemiology, Oregon Health & Science University, Portland, OR, USA.
5. Department of Biomedical Engineering, University of Florida, Gainesville, FL, USA
6. UCSD Health Department of Biomedical Informatics, UCSD, La Jolla, CA 92093, USA.
7. Department of Informatics Laboratory Medicine, Informatics Section, Yale SoM, New Haven, CT 06520, USA
8. Department of Medicine, Division of Genomics and Precision Medicine, UCSD, La Jolla, CA 92093, USA.
9. Department of Otolaryngology, USF Health Voice Center, University of South Florida, Tampa, FL, USA.
10. Department of Biomedical Informatics at Vanderbilt University, Nashville, TN, USA.

* Authors contributed equally.

# Joint corresponding authors: Peipei Ping and John Rincon.

## ABSTRACT

**Objective:**

As AI becomes increasingly central to healthcare, there is a pressing need for bioinformatics and biomedical training systems that are personalized and adaptable.

**Materials and Methods:**

The NIH Bridge2AI Training, Recruitment, and Mentoring (TRM) Working Group developed a cross-disciplinary curriculum grounded in collaborative innovation, ethical data stewardship, and professional development within an adapted Learning Health System (LHS) framework.

**Results:**

The curriculum integrates foundational AI modules, real-world projects, and a structured mentee-mentor network spanning Bridge2AI Grand Challenges and the Bridge Center. Guided by six learner personas, the program tailors educational pathways to individual needs while supporting scalability.

**Discussion:**

Iterative refinement driven by continuous feedback ensures that content remains responsive to learner progress and emerging trends.

**Conclusion:**

With over 30 scholars and 100 mentors engaged across North America, the TRM model demonstrates how adaptive, persona-informed training can build interdisciplinary competencies and foster an integrative, ethically grounded AI education in biomedical contexts.

## 1. INTRODUCTION

**1.a. Problem Background.** AI's transformative potential in healthcare and biomedicine has outpaced current training and skill acquisition, revealing critical gaps across the field. Present challenges include limited educational accessibility, partiality and reproducibility issues, insufficient integration with clinical practice, and a lack of sustained engagement with real-world stakeholders. Compounding these is the urgent need for curricula that can evolve in step with AI's rapid advancement. The Bridge2AI TRM-WG curriculum addresses these deficits through a flexible, accessible educational model that fosters cross-disciplinary collaboration, prioritizes ethically sourced data, and cultivates professional skills promoting accountability, adaptability, and meaningful application in biomedical and healthcare settings.

**1.b. Infrastructure.** The Bridge2AI Consortium comprises the Bridge Center (BC) and Grand Challenges (GC). The BC integrates people, teams, programs, and activities across six components focused on various aspects of the biomedical AI ecosystem and training. Full descriptions of the BC Working Groups (WGs) are provided in **Supplementary Table 1**. The GCs are a series of ambitious data generation projects aimed at harnessing the potential of AI across varied domains of health research and detailed in **Table 1**. Each GC is uniquely designed to promote the integration of AI in healthcare and public health by developing ethically-sourced (i.e., following HIPAA and IACIC guidelines) and AI-ready datasets.

## 2. METHODS

**2.a. Context.** The TRM-WG was charged with developing a dynamic, adaptive framework to meet the evolving demands of AI technologies and their translational applications in clinical contexts. This framework addresses three critical educational pillars, detailed within the **Definitions** section of the **Supplementary Materials**: (I) cross-disciplinary collaborative innovation; (II) data stewardship and consideration; and (III) professional development through applied skills.

**2.b. Intervention - Creating a Cross Disciplinary Framework.** The modular TRM-WG training program addresses the above needs and is broadly accessible to varied areas of professional expertise. The curriculum leverages a library of training resources across the consortium, spanning multiple foundational modules and dataset-specific training materials. This framework adapts elements across three models: TSBM[1], Plan-Do-Study-Act[2], and the LHS[3]. The LHS acts as core scaffolding rooted in principles of evidence-based practice, continuous learning, and collaboration among healthcare stakeholders, adapting six key components (**Figure 1A**). LHS also broadly emphasizes learning cycles which are adapted throughout the consortium with a trickle-down effect to shape individual curricula for mentees.

**2.c. Intervention - Defining the Bridge2AI Scholars.** Bridge2AI Scholars are consortium-wide investigators, learners, mentors, and mentees. Mentees are motivated individuals eager to advance their AI education in biomedical sciences, as exceptional candidates recruited across the consortium. Mentors are experts in their respective fields and are tasked with guiding mentees through their personalized curriculum action plan (**Figure 1B**). These investigators are knowledgeable in their respective GC's AI data generation and use protocols, as well as domain AI applications. Mentees match with multiple Bridge2AI mentors based on their background, interests, and research goals, comprising a multidisciplinary mentor team with expertise from clinical, computational, and AI leadership backgrounds. This ensures comprehensive guidance across the cross-disciplinary complexities of AI in health sciences. This tailored mentee-mentor approach guides mentees to address specific use cases (e.g., drug response predictions, medical imaging analysis, and sentiment analysis) to deepen the mentees' understanding of real-world datasets and applications relevant to their research interests. This mentee-mentor networking is designed to nurture continuous professional development and a culture of collaboration and cross-disciplinary innovation.

**2.d. Intervention - Cultivating and Maintaining the Scholar Network and Portal.** he Scholars Storyboard Padlet, a public online blog-style platform that allows Bridge2AI scholars, was implemented as part of the program efforts to build a strong and collaborative Bridge2AI Scholar Network. On this platform, both mentors and mentees introduce themselves to the community, detailing interests and experience. This tool provides scholars with a structured yet user-friendly way to view a roster of fellow Bridge2AI scholars. Padlet helps to inspire and kickstart networking opportunities to engage in cross-disciplinary collaboration.

**2.e. Design and Analysis.** The BC integrates the LHS framework within TRM-WG operations to create a dynamic and responsive ecosystem, continuously refining and optimizing curricular offerings based on feedback from stakeholders. Feedback is collected from all stakeholders in an ongoing process, informing the curricular improvement and annual program evaluation. This program addresses the following educational needs: cross-disciplinary collaboration, data stewardship, professional development, and applied skills. Experts identify domain-specific challenges and solutions via the BC's collaborative efforts. Activities in data sharing and dissemination promote standardized data collection practices and knowledge generation. This multifaceted alignment with the LHS framework enhances the ability to identify problems, integrate varied educational resources, engage stakeholders meaningfully, and enable iterative refinement of practices and protocols.

## 3. RESULTS

**3.a. Outcome - Successful Deployment of the Cross Disciplinary Framework.** By adapting the LHS model, this curriculum framework meets two objectives: (1) creating a flexible and adaptable training curriculum that bridges the educational gaps in biomedical AI for a wide range of learners, and (2) providing targeted training tailored to the specific challenges and opportunities of each partner GC. The TRM-WG's address skill gaps in AI competencies and foster cross-disciplinary collaboration among 150+ scholars (30+ mentees and 100+ mentors). Outcome data show positive shifts in knowledge, practical skills, and ethical awareness. Ongoing analysis reveals the curriculum's capacity to adapt effectively, with learners reporting increased confidence in understanding and applying AI within biomedical research.

**3.b. Outcome - Identified Relevant Domain Specific Personas.** The Bridge2AI TRM-WG identified six personas as categories representing the learners largely encountered in health data science[4–7]. These consist of: Healthcare Professionals, Clinical Research Scientists, Bioinformaticians, Clinical Systems Engineers, Behavioral Data Scientists, and Public Health Analysts. Each persona is assessed across six core areas relevant to biomedical AI: Clinical Knowledge & Applications, Molecular Biology & Omics Data, Bioethics & Legal Standards, Data Science & Statistics, AI Algorithms & Theory, and Programing & Computation (see **Supplementary Materials Definitions**). **Figure 2** illustrates radar charts to visualize levels of mastery and identify learning opportunities across each persona to identify a hypothetical individual's initial mastery and target growth.

**3.c. Outcome - Personalized Curriculum and Dynamic Mentor Support.** To address diverse skill gaps, the TRM-WG categorizes learners into distinct personas and develops targeted, individualized curriculum action plans. During onboarding, mentees collaborate with their tailored mentor teams to define learning profiles and co-create flexible plans, progressing through ESAI lectures, AI computation, theory, and biomedical applications at their own pace. These scaffolded plans include real-world projects, iterative feedback, and mastery tracking, aligning with LHS principles of adaptive learning (**Figure 1B**).

Mentors selected for their interdisciplinary expertise guide mentees through ongoing support. Educational interventions enabled by the mentor-mentee model reinforce continual curriculum development. By integrating varied perspectives, the curriculum cultivates innovative thinking and prepares scholars to navigate biomedical AI challenges. The scholar network facilitates resource-sharing to improve curriculum outreach and collaboration (**Supplementary Figure 1** and **Figure 3**).

**3.e Outcome - Improved Translational AI Biomedical Learning.** Success stories exemplify the program's adaptability and its impact on scholars across varied health domains, advancing AI integration in health sciences education and promoting meaningful cross-collaboration (**Supplementary Materials)**.

**3.f. Contextual Factor - Consortium-wide Collaboration.** Operating under a shared governance model, the Bridge2AI program emphasizes transparency, partnership, and accountability. This governance structure supports widespread engagement with various stakeholders, including other research programs, to leverage additional resources and amplify the program's impact **(Figure 1B)**. Each phase of the program is implemented and transformed through regular feedback loops and collaboration with different WG's. For example, the TRM-WG's educational efforts are reinforced by extensive support from BC and GC partnerships **(Supplementary Table 1)**. They provide unique learning opportunities, with tailored resources and real-world datasets enhancing learner engagement and enabling a deeper connection to AI in biomedical contexts. Each GC incorporates real-world data problems that reinforce the LHS approach, promoting an iterative process where curriculum evolves through learner feedback and cross-group collaboration. These integrated efforts establish a comprehensive governance and operational framework for the Bridge2AI consortium as it pertains to education, ensuring that AI protocols are responsibly developed and applied, adhering to the highest standards of scientific and ethical rigor. Cross-collaboration and dissemination of educational content is enhanced, and continuous real-time improvement of training materials is augmented with feedback from consortium members represented in the TRM-WG. This dynamic resource-sharing ecosystem is critical to the LHS framework, promoting the practice of collective learning and adaptation, ensuring that resources are effectively utilized and refined to meet the evolving needs of biomedical AI education. The resources offered by the consortium engages a broad spectrum of stakeholders.

**3.g Contextual Factor - Consortium-Wide Learning Resources and Activities.** Bridge2AI offers a rich ecosystem of ethically grounded, scientifically robust educational resources developed collaboratively by all partner GCs. The TRM-WG curates training materials spanning foundational AI concepts, ethics, and real-world applications, integrating case studies that align with LHS principles to bridge theory and practice. Each GC delivers domain-specific learning activities tailored to its biomedical focus. These include 12-month training programs, datathons, micro-learning sessions, internships, and co-internship opportunities, promoting ethical data stewardship and community engagement. Initiatives such as public webinars, hackathons, and outreach campaigns foster professional development and skill-building for varied career paths. Bi-annual Face-to-Face meetings and Open Houses hosted by partner institutions highlight scholar achievements, encourage peer learning, and build a vibrant, interdisciplinary community. Monthly TRM-WG meetings ensure continuous curricular refinement through collaboration and shared governance. (**Table 1** and **Table 2**).

## 4. CONCLUSIONS

The TRM-WG has laid the foundation for a training curriculum aligned with the needs of the Bridge2AI consortium's learners, enhancing the effectiveness of the programs and equipping a versatile workforce to tackle challenges at the intersection of AI and healthcare. This dynamic, adaptive curriculum integrates ethical considerations, continuous mentorship, and a cross-disciplinary framework. By addressing educational gaps and fostering cross-disciplinary collaboration, the curriculum effectively supports varied learner personas within the biomedical enterprise. This adaptive approach continues to evolve, ensuring relevance and impact while promoting innovation in biomedical AI training. These efforts exemplify a model of integration and adaptability that can inspire similar consortia globally, ultimately contributing to improved health outcomes for all

## 5. FUTURE DIRECTIONS

**5.a Bridge2AI Scholars Network Refinement and Expansion.** Establishing a cohesive central curriculum is both formidable and unsuitable for the rapidly evolving field. Some practices in traditional standardized curricula may inadvertently limit access in certain situations. Acknowledging the cross-disciplinary nature of AI biomedical research and AI-enhanced clinical applications reveals a need for adaptive foundational curricula open to iterative improvement to support the entire learning community. The TRM-WG are committed to this adaptive strategy.

**5.b. Fostering Mentee-Mentor Collaborations.** Each GC offers distinct learning activities at different times throughout the year. This opportunity enables the TRM-WG to embed learning activities from each GC throughout a learner's journey in a staggered fashion, which requires consistent communication. Future improvements include improving and maintaining a committed and consistent line of communication with all Bridge2AI Scholar Program partners, mentees and mentors, to ensure timely updates, effective mentorship, and mentee growth. Collaborations with the TC-WG are planned to refine the Bridge2AI Scholar Program and standardize a protocol for connecting learners with mentors, streamlining the Bridge2AI Scholar pipeline. Working closely with other WGs in the Bridge2AI consortium will enhance the consistency and effectiveness of mentorship connections, fostering a more robust and supportive educational environment for all Bridge2AI scholars.

**5.c Enriching Ethics and Social Awareness in AI Education.** The TRM-WG also anticipates collaborating closely with the ELSI-WG to enhance the ethical framework, standards, and best practices embedded in the Bridge2AI curricula. This collaboration will support regular updates to these educational resources, ensuring they reflect the latest advancements in ethical AI. Areas for expansion include helping learners critically examine the provenance of data—asking whether it is ethically sourced, whether collection methods are transparent, and whether accountability mechanisms exist for data collection, storage, and handling. Learners will be guided to address ethical dilemmas and ensure data is used solely within its intended scope. The ELSI-WG's expertise will be essential in shaping these discussions and enriching the ethical depth of the training materials.

## ACKNOWLEDGEMENTS

The content of this manuscript does not necessarily represent the official views of the National Institutes of Health, the Department of Health and Human Services, or the United States Government. The authors are grateful for all members of the Bridge2AI consortium, their generous support, and collaborative efforts.

## COMPETING INTERESTS

The authors have no competing interests to declare.

## FUNDING

This work was supported by the following grants and institutions for their support and contributions to the Bridge2AI program:

- **Building BRIDGEs** (Project Number: 1U54HG012517-01), awarded to the University of California, Los Angeles.
- **Integration, Dissemination and Evaluation (BRIDGE) Center for the NIH Bridge to Artificial Intelligence (BRIDGE2AI) Program** (Project Number: 1U54HG012513-01), awarded to the University of Colorado Denver.
- **A FAIR Bridge2AI Center (FABRIC)** (Project Number: 1U54HG012510-01), awarded to the University of California, San Diego.
- **Bridge2AI: Voice as a Biomarker of Health - Building an Ethically Sourced, Bioacoustic Database to Understand Disease Like Never Before** (Project Number: 1OT2OD032720-01), awarded to the University of South Florida.
- **Bridge2AI: Cell Maps for AI (CM4AI) Data Generation Project** (Project Number: 1OT2OD032742-01), awarded to the University of California, San Diego.
- **Bridge2AI: Patient-Focused Collaborative Hospital Repository Uniting Standards (CHoRUS) for Equitable AI** (Project Number: 1OT2OD032701-01), awarded to Massachusetts General Hospital.

## REFERENCES

1. Luke DA, Sarli CC, Suiter AM, et al. The Translational Science Benefits Model: A New Framework for Assessing the Health and Societal Benefits of Clinical and Translational Sciences. *Clinical Translational Sci*. 2018;11(1):77-84. doi:10.1111/cts.12495
2. Taylor MJ, McNicholas C, Nicolay C, Darzi A, Bell D, Reed JE. Systematic review of the application of the plan-do-study-act method to improve quality in healthcare. *BMJ Qual Saf*. 2014;23(4):290-298. doi:10.1136/bmjqs-2013-001862
3. Friedman C, Rubin J, Brown J, et al. Toward a science of learning systems: a research agenda for the high-functioning Learning Health System. J Am Med Inform Assoc. 2015;22(1):43-50. doi:10.1136/amiajnl-2014-002977
4. Wood EA, Ange BL, Miller DD. Are We Ready to Integrate Artificial Intelligence Literacy into Medical School Curriculum: Students and Faculty Survey. *Journal of Medical Education and Curricular Development*. 2021;8:238212052110240. doi:10.1177/23821205211024078
5. Treuillier C, Boyer A. Identification of class-representative learner personas. In: ; 2021. https://api.semanticscholar.org/CorpusID:247625913
6. Yang X. Creating learning personas for collaborative learning in higher education: A Q methodology approach. *International Journal of Educational Research Open*. 2023;4:100250. doi:10.1016/j.ijedro.2023.100250
7. Mulder N, Schwartz R, Brazas MD, et al. The development and application of bioinformatics core competencies to improve bioinformatics training and education. Troyanskaya OG, ed. *PLoS Comput Biol*. 2018;14(2):e1005772. doi:10.1371/journal.pcbi.1005772

## FIGURES

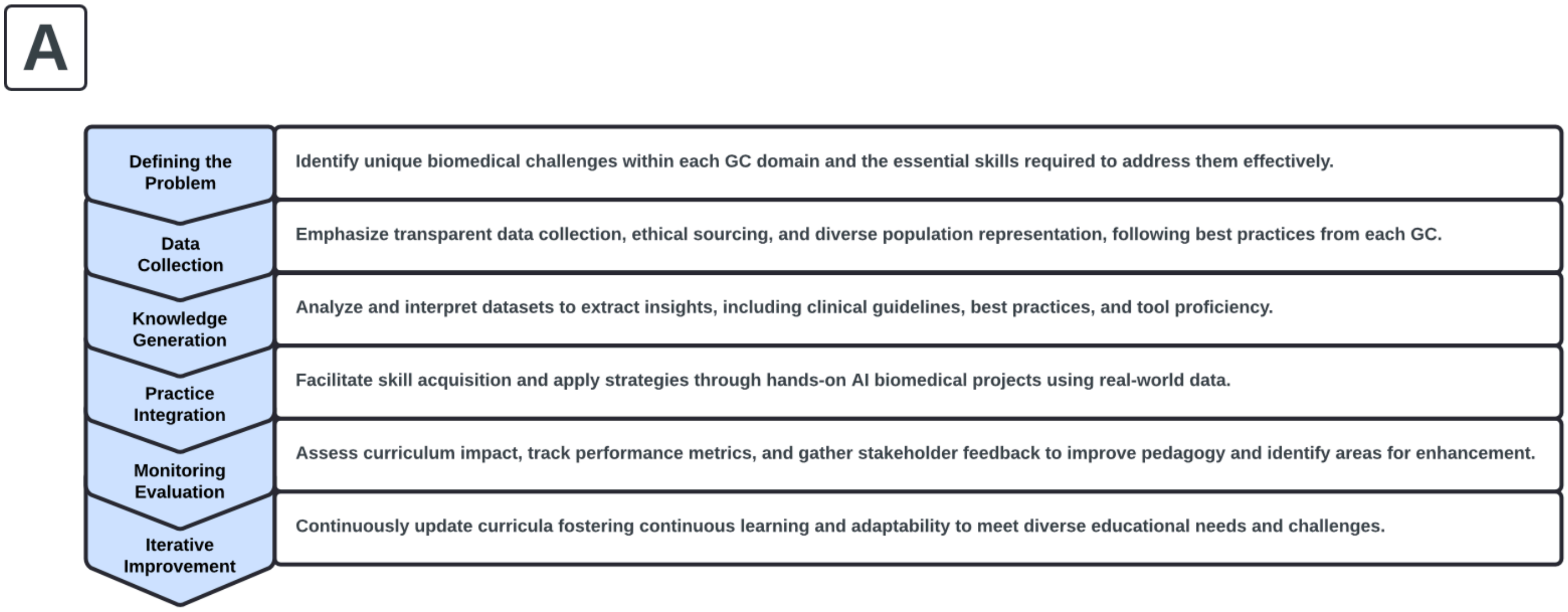

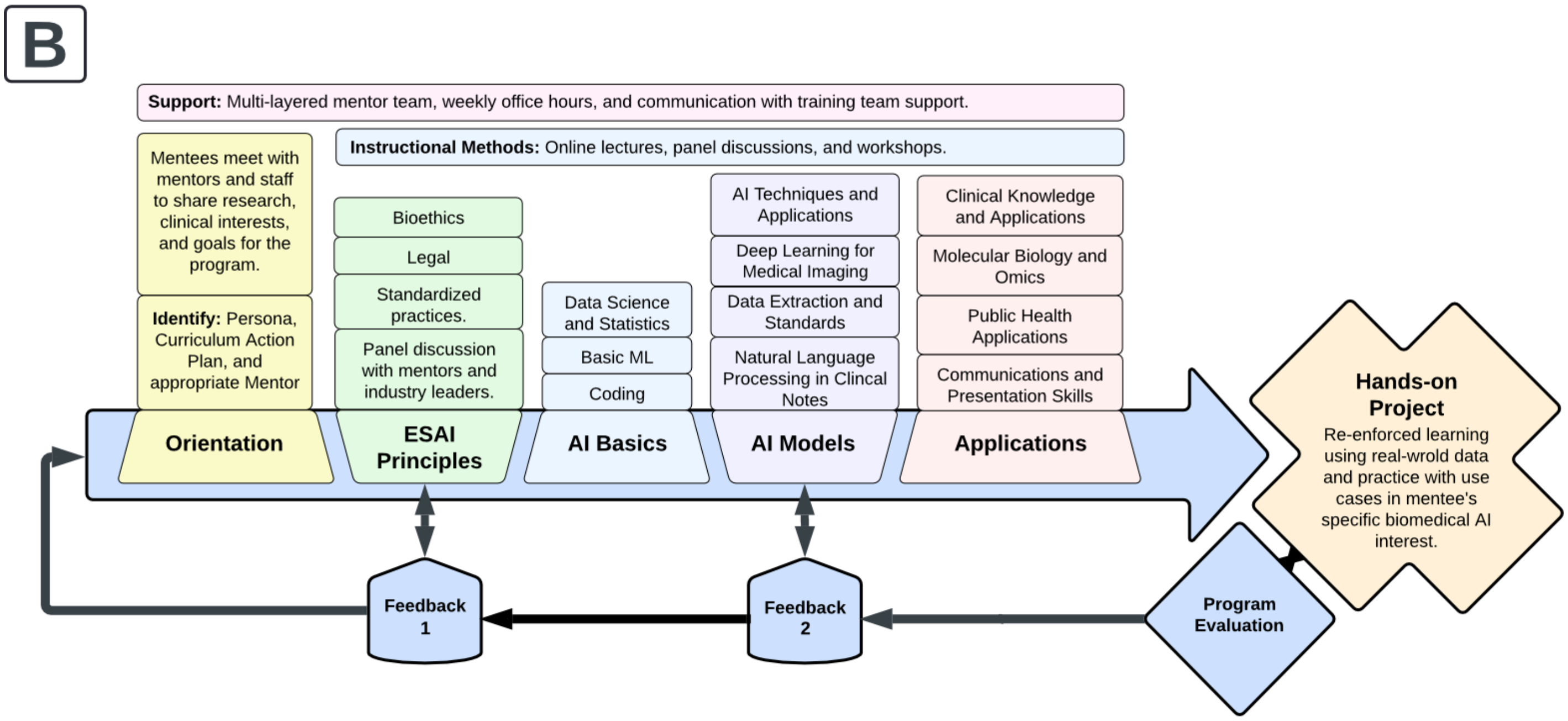

**Figure 1. Cross-Disciplinary Curriculum Framework and Personalized Action Plan.**

Together, the curriculum development framework and personalized action plan establish a comprehensive, adaptive training model that integrates cross-disciplinary design with individualized learning pathways. *A)* Rooted in the Learning Health System (LHS) model, this curriculum development framework emphasizes evidence-based practice, continuous learning, and collaboration. It consists of six interlocking components designed to support the co-creation of a dynamic, cross-disciplinary curriculum. *B)* The Personalized Curriculum

Action Plan. This process begins with an onboarding interview to assess current expertise and learning goals to determine a tailored learning pathway. All mentees engage with core Ethics and Social Awareness in AI lectures, advance through varied AI competencies in biomedicine, and apply knowledge through real-world projects. Ongoing feedback mechanisms enable continuous refinement of both foundational and advanced content.

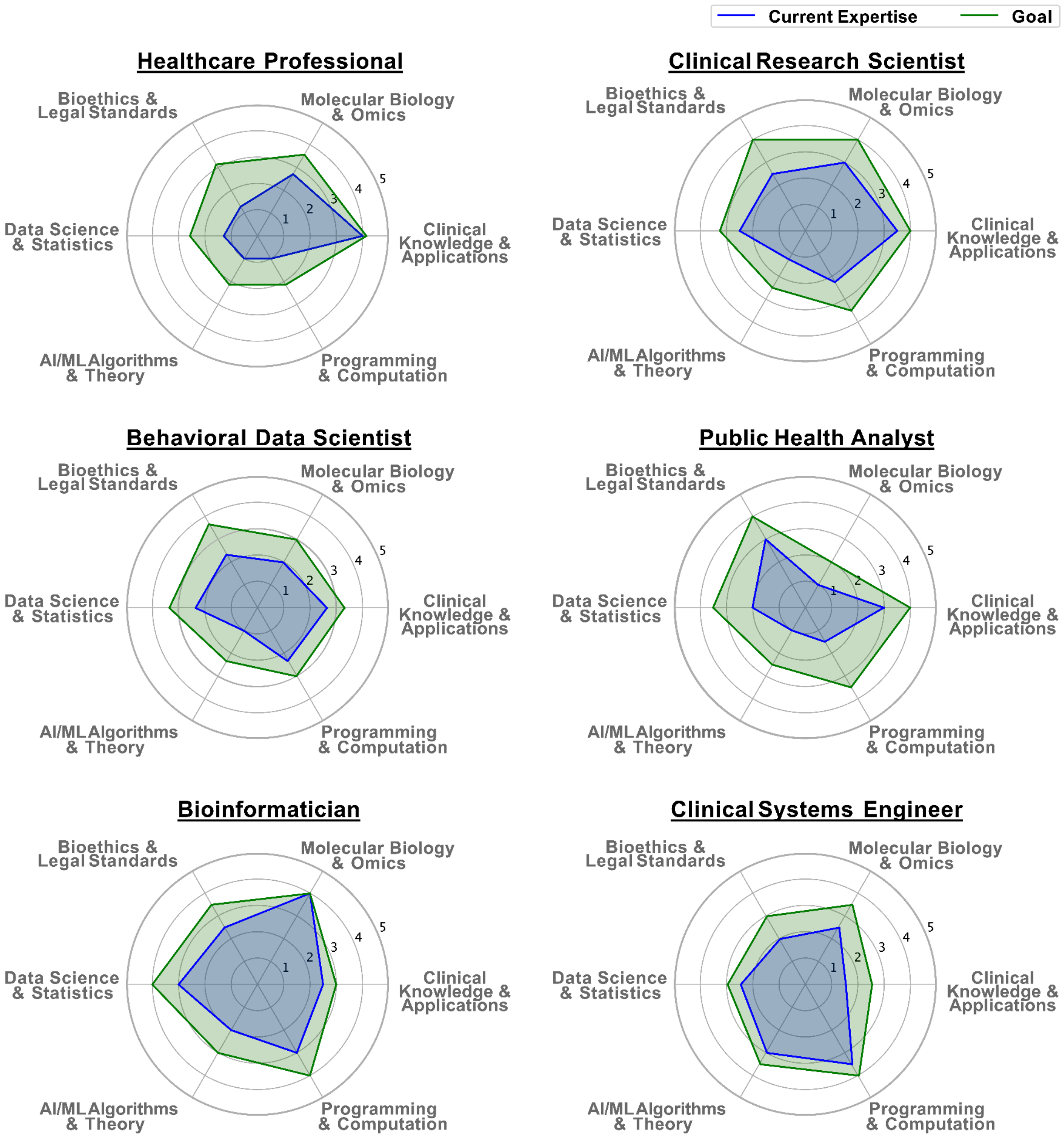

**Figure 2. Learner Personas.** These radar charts illustrate characteristic profiles for each learner persona

across six key domains: data science & statistics, AI algorithms & theory, programming & computation, clinical knowledge & applications, and molecular biology & omics. The blue values reflect estimated baseline exposure, inferred from each persona’s typical academic or professional background and the standard competencies associated with those fields. The green values represent anticipated growth following completion of the personalized curriculum action plan, informed by the educational modules delivered throughout the program. This approach provides a conceptual visualization of each persona’s learning journey, offering a framework to guide curriculum development and alignment with core training goals.

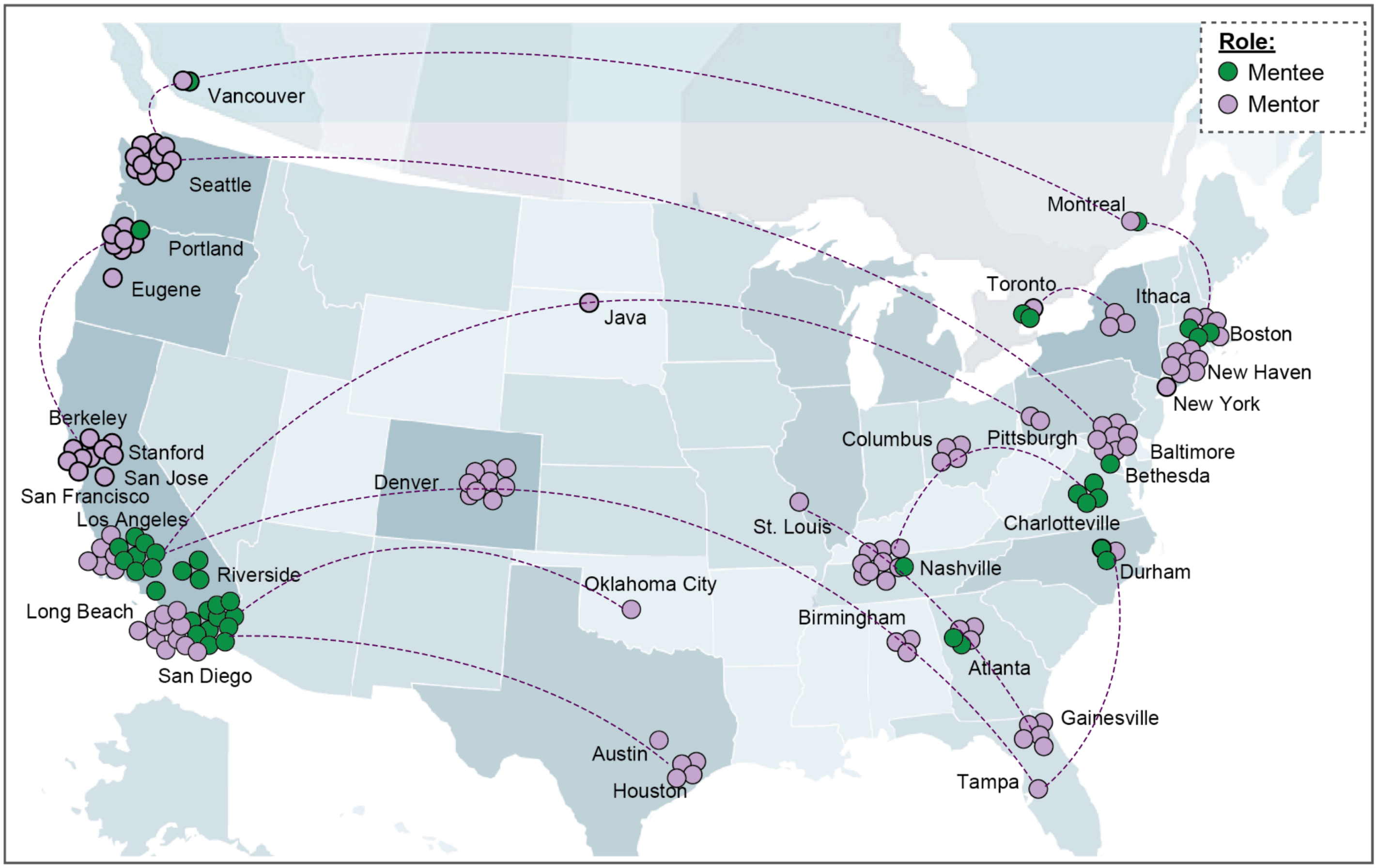

**Figure 3. Geographic Distribution of Bridge2AI Scholars and Mentors Engaged in B2AI Program.** The Bridge2AI scholars and mentors are from both academic institutions and various sectors of healthcare that span across 34 cities throughout 20 US states and 3 cities in Canada. Dotted lines demonstrate a few examples of consortium wide network collaborations which overcome geographic constraints, purple dots representing mentors and mentees in green.

**TABLES**

| Bridge2AI Grand Challenge (GC) | Details |
|---|---|
| Precision Public Health / Voice (GC-Voice) | The Voice GC integrates voice as a biomarker in clinical care by establishing a varied, multi-organizational voice database leveraging data from vocal pathologies, neurological disorders, and various conditions to develop predictive models for a broad range of diseases. Learn more about this GC at this website, (https://www.b2ai-voice.org/). |
| Functional Genomics / CM4AI (GC-CM4AI) | The Cell Maps for AI (CM4AI) GC utilizes advanced mapping techniques such as proteomic mass spectrometry, cellular imaging, and CRISPR/Cas9 genetic perturbation to map the spatiotemporal architecture of human cells and facilitate interpretable learning of genotype-phenotype relationships. Learn more about this GC at this website, (https://cm4ai.org/) |
| Clinical Care / CHoRUS (GC-CHoRUS) | The AI for Clinical Care Network (CHoRUS) harmonizes multimodal EHR, waveform, imaging, and text data while incorporating Social Determinants of Health into recovery models in an effort to enhance recovery from acute illnesses. Learn more about this GC at this website, (https://chorus4ai.org/) |
| Salutogenesis / AI-READI (GC-AI-READI) | The Artificial Intelligence Ready and Equitable Atlas for Diabetes Insights (AI-READI) project focuses on creating a flagship dataset to serve as a basis to map the temporal atlas of type 2 diabetes development and its potential reversal, providing invaluable |
| | insights into the salutogenic processes. Learn more about this GC at this website, (https://aireadi.org/). |

**Table 1. Bridge2AI Grand Challenges Overview.** This table provides an overview of the Bridge2AI Grand Challenges (GCs) that directly inform and enrich the curriculum. Each GC represents a unique biomedical AI domain, offering distinct datasets, methodologies, and learning opportunities that support the design of practical, data-driven training modules. Including this table underscores the collaborative and cross-disciplinary nature of the curriculum's foundation.

| TRM Learning Resource Modules | Details |
|---|---|
| **Ethics and Social Awareness in AI (ESAI)** | This module features lectures promoting the understanding and implementation of ethical AI practices in health science contexts. It covers principles like data privacy, data integrity, and the responsible use of AI in health applications (https://vimeo.com/showcase/10029782). |
| **The Fundamental Concepts of AI** | This module features lectures carefully organized to introduce foundational concepts and state of the art technologies supporting AI, including basic elements of AI, common practices in data science, and advanced AI models/applications (https://vimeo.com/showcase/10035538). |
| **Datasets in Grand Challenges** | This module provides resources for users to learn about each GC's respective datasets, including standards and analysis tools. Potential research opportunities or challenges are also detailed. Additional lectures will feature hands-on tutorials and use-case examples specific to each GC (https://vimeo.com/showcase/10035544). |
| **Special Topic Webinars** | This module includes lectures on emerging technologies relevant to AI, government policies, and funding opportunities. Topics so far include large language models, AI in clinical decision-making, and integrating AI tools into existing healthcare systems (https://vimeo.com/showcase/10035547). |

**Table 2. Accessible Online TRM Resources Overview.** This table outlines key online learning modules developed by the TRM-WG in partnership with Bridge2AI Grand Challenges and the Bridge Center. These curated resources provide scalable, foundational-to-advanced content in biomedical AI, ethics, core AI concepts, data use, and emerging technologies.

## Supplementary Materials

The TRM-WG offers educational material which is broadly accessible to professionals across many domains and specialties. These contents are organized as a consortium-wide library of training resources that includes both core modules and materials tailored to specific datasets **Supplementary Figure 2**. This shared network consolidates and amplifies the educational efforts across the Grand Challenges (GCs) as well as the Bridge Center (BC). The GCs develop dataset- and domain-specific educational materials. The Working Groups (WGs) of the BC focus on consortium-wide functions (**Supplementary Table 1**).

To support the educational goals of the scholars, the TRM-WG categorizes learners into distinct personas (**Figure 2**). This helps the mentee-mentor team develop a targeted curriculum action plan that addresses specific skill gaps and learning needs, enabling a personalized approach to education (**Figure 1B**). Many learners fit into multiple personas. This is addressed by matching their educational needs by connecting them with appropriate mentors relevant to their personas. **Supplementary Figure 1** illustrates tailored educational trajectories aimed at achieving persona-specific target growth.

**Definitions:**

Three critical educational pillars:

I. **Cross-disciplinary Collaborative Innovation**: Educational strategies that prioritize human-centered AI through collaboration across disciplines, aiming to identify ethically sound and clinically meaningful AI applications throughout the healthcare ecosystem.

II. **Data Stewardship & Considerations**: Curricula designed to instill responsible data practices, emphasizing the ethical sourcing of data, fairness, accurate representation of patient populations, and the integrity of metadata—aligned with NIH ACD AI Working Group recommendations.[1]

III. **Professional Development and Applied Skills**: Ongoing learning initiatives that integrate real-world

[1] Report of the ACD AI Working Group. Published online December 6, 2019. https://acd.od.nih.gov/working-groups/ai.html

use cases, cultivate nuanced understanding, and promote mentorship structures to support emerging professionals in biomedical AI.

Six foundational knowledge areas integrated into the Bridge2AI curriculum:

1. **Clinical Knowledge & Applications:** Comprehensive understanding and practical use of medical knowledge in patient care, including expertise in diagnosis, treatment, patient management, and effective communication while incorporating ethical, legal, and technological aspects to enhance healthcare outcomes.
2. **Molecular Biology & Omics Data:** Understanding the structure, function, and interactions of biomolecules, as well as utilizing omics technologies to analyze and interpret complex biological data for advancements in research, diagnostics, and therapeutics.
3. **Bioethics & Legal Standards:** Expertise of ethical principles and legal frameworks within healthcare and biomedical research; it encompasses analyzing and addressing moral dilemmas, establishing guiding practices within biomedical enterprise, and ensuring compliance with laws and regulations.
4. **Data Science & Statistics**: The understanding of statistical analysis, machine learning, data mining, and visualization techniques to identify patterns, make predictions, and drive evidence-based practices.
5. **AI Algorithms & Theory:** Knowledge of the mathematical and conceptual foundations that underpin the design, analysis, and optimization of AI algorithms, including linear algebra, calculus, probability, statistics, machine learning theory, and neural networks.
6. **Programming & Computation:** Proficiency in programming languages and computational skills. Emphasis on practical application, software engineering, development workflows, and use of development tools to create, optimize, and maintain software applications and systems.

**Success Training Outcomes:** these anonymized examples showcase the TRM-WG curriculum's successful training outcomes:

★ Cardiovascular Health: A Bridge2AI scholar gained hands-on experience with clinical machine learning, contributing valuable insights into cardiovascular diseases and expanding their roles in clinical

healthcare applications.

- ★ Oncology Intern: One scholar focusing on cancer immunotherapy has gained experience applying AI to clinical studies, advancing new analytical approaches for the understanding of cellular interactions in cancer treatment.
- ★ Behavioral Science: A developmental psychology scholar acquired new expertise in AI-driven predictive modeling, now adapting expansive data practices and community engagement in their work.
- ★ Healthcare Practice: A clinician specializing in ocular pathology expanded their analytical capabilities in AI-enabled image analyses, leveraging collaborative opportunities from their mentor team.
- ★ Bioethics: An investigator interested in bioethics with a foundation in life sciences found the program instrumental in applying AI concepts to interdisciplinary research, enhancing their work alongside clinical researchers.

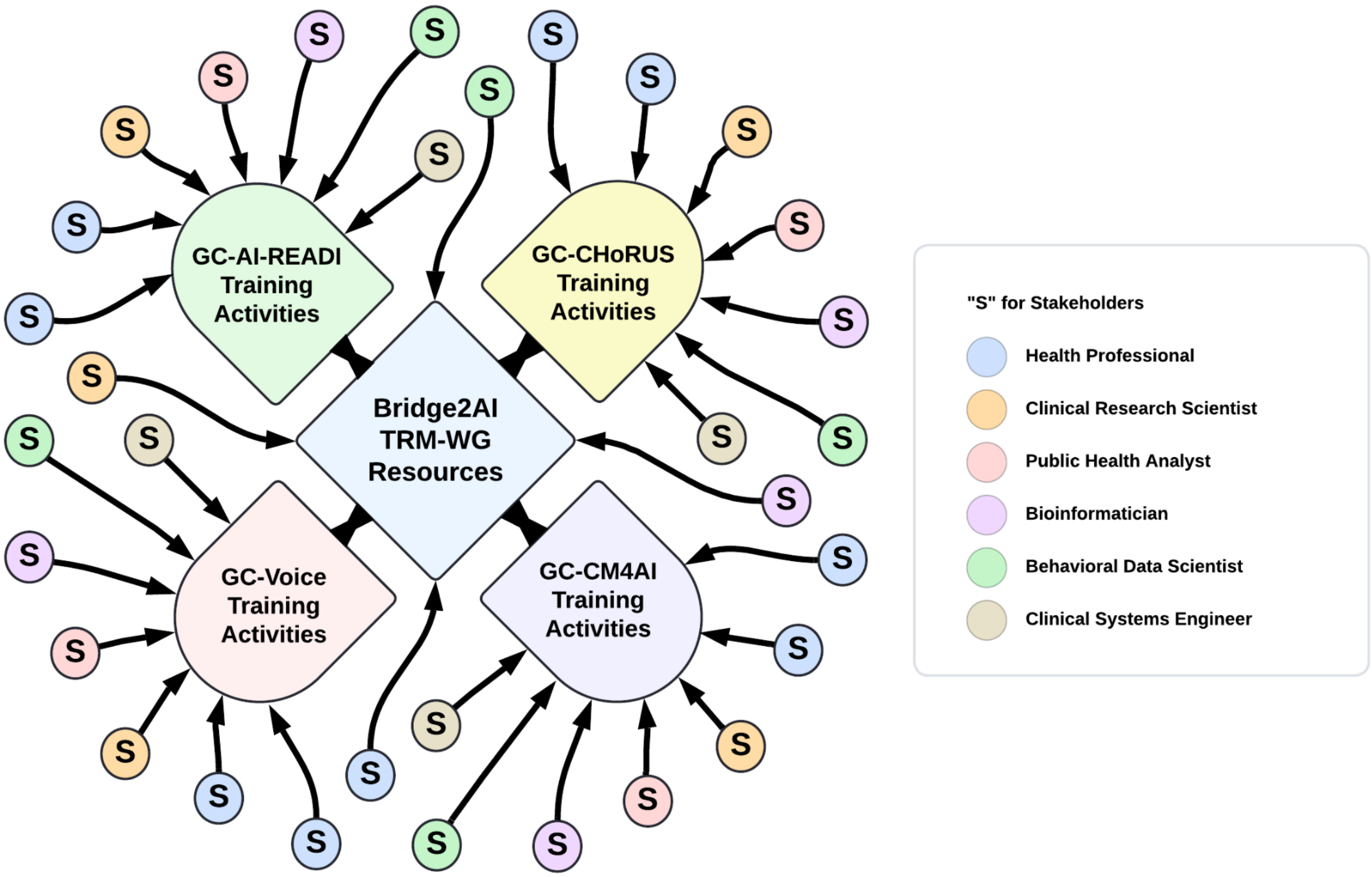

**Supplementary Figure 1. The Bridge2AI TRM-WG Network.** Visualization of the Bridge2AI TRM-WG and partner GCs resource-sharing network to improve curriculum outreach for AI biomedical training, investigation, and innovation. Prospective Bridge2AI learners are denoted with an “S” for stakeholder, their interest in specific AI biomedical fields is denoted with different colors, and arrows illustrate the consortium’s collective outreach efforts.

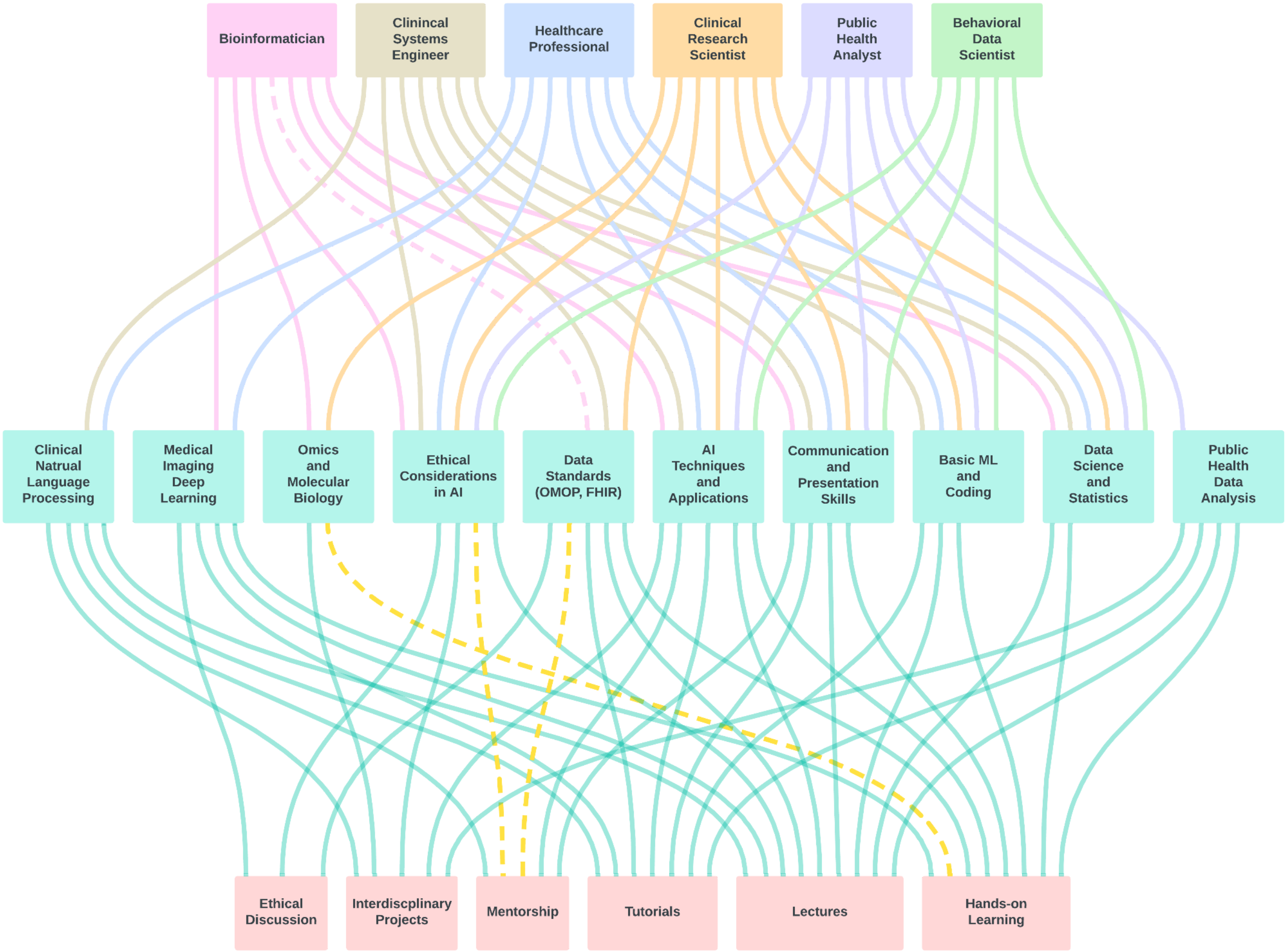

**Supplementary Figure 2. Persona-based Curricula.** This Sankey Diagram illustrates the interdependencies between learner personas, areas of study, and the instructional mediums through which these topics are navigated. The selected areas of study serve as a baseline but are not limited to, as the skills and competencies necessary for each learner persona to succeed in their specialization vary. Green edges between areas of study and instructional mediums indicate content available at the time of publication, while proposed content is indicated in dashed gold nexus.

| Bridge2AI Working Group | Role / Details |
|---|---|
| Standards, Practices, and Quality Assessment (SPQ-WG) | Identify community needs, coordinate standards implementation, and enable sharing and reuse by promoting open-source collaboration, transparent benchmarking, and best practices for efficient data dissemination. |
| Training, Recruitment, and Mentoring (TRM-WG) | Develops educational materials that blend technical skills with awareness of socio-technical impacts to prepare a varied cohort of future scientists for success in AI-enhanced biomedicine. |
| Teaming and Collaboration (TC-WG) | Promotes cross-disciplinary collaboration using evidence-based team science principles to enhance consortium cohesion and operational synergy through structured frameworks and toolkits. |
| Data Sharing and Dissemination (DSD-WG) | Explores and enforces ethical standards for sharing and disseminating open biomedical data to support data-sharing initiatives by offering guidelines and resources to uphold high standards of openness and ethical conduct in data management. |
| Data Sharing and Open House (DSOH-WG) | Outlines protocols for showcasing consortium-wide achievements to the public and engages external stakeholders through annual events, cultivating interaction with broader scientific and public communities. |
| Communications, Meetings, and Portal (CMP-WG) | Centralizes consortium communications and meeting logistics, ensuring consistency, compliance, and transparency. |
| Ethical, Legal, and Social Implications (ELSI-WG) | Develops ethical guidelines for AI in biomedical settings, addressing issues like data ethics, consent, and partiality. |

**Supplementary Table 1. Bridge2AI Working Group Overview.** This table provides an overview of the seven key working groups (WGs) within the Bridge2AI consortium, detailing their primary functions and contributions. The left column identifies each working group by name, while the right column describes the specific roles and objectives of each group in advancing ethical, operational, and educational standards in AI-enhanced biomedicine.